\documentclass[aps,showpacs,prl,twocolumn,letterpaper]{revtex4-1}

\pdfoutput=1

\usepackage{graphicx}
\usepackage{amsmath}
\usepackage{color}

\begin{document}

\title{Traveling and resting crystals in active systems}

\date{\today}

\author{Andreas M.~Menzel}
\email[email: ]{menzel@thphy.uni-duesseldorf.de}
\affiliation{Institut f\"ur Theoretische Physik II: Weiche Materie, Heinrich-Heine-Universit\"at D\"usseldorf, Universit\"atsstra{\ss}e 1, D-40225 D\"usseldorf, Germany}
\author{Hartmut L\"owen}
\affiliation{Institut f\"ur Theoretische Physik II: Weiche Materie, Heinrich-Heine-Universit\"at D\"usseldorf, Universit\"atsstra{\ss}e 1, D-40225 D\"usseldorf, Germany}

\begin{abstract}
A microscopic field theory 
for crystallization in active systems is proposed which unifies
the  phase-field-crystal model of freezing with the Toner-Tu theory 
for self-propelled particles. A wealth of different active crystalline 
states are predicted and characterized. In particular, for increasing 
strength of self-propulsion, a transition from a resting crystal to a 
traveling crystalline state is found where the particles 
migrate collectively while keeping their crystalline order. 
Our predictions, which are verifiable in experiments  and in particle-resolved 
computer simulations, provide a starting point for the design of new active materials.
\end{abstract}


\maketitle

Self-propelled particles \cite{vicsek1995novel} exhibit fascinating collective phenomena  like 
swarming, swirling and laning which have been intensely explored by 
theory, simulation and experiment, for recent reviews 
see \cite{ramaswamy2010mechanics,romanczuk2012active,cates2012diffusive}. 
In marked contrast to passive particles,
self-propelled ``active'' particles 
have an internal motor of propulsion, dissipate 
energy and are therefore intrinsically in nonequilibrium.
Examples of active particles 
include living systems, like bacteria and microbes \cite{drescher2011fluid}, 
as well as man-made microswimmers, catalytically driven 
colloids \cite{paxton2004catalytic,volpe2011microswimmers} and granular hoppers \cite{narayan2007long}. 

If, at high densities, the particle interaction dominates the propulsion,
crystallization in an active system is conceivable. It is 
expected that such ``active crystals'' have structural and dynamical 
properties largely different from equilibrium crystals due to the intrinsic
drive. In fact, there is  experimental evidence for
active crystals, both from observations of hexagonal 
structures for catalytically-driven colloids
 \cite{theurkauff2012dynamic} and honeycomb-like textures for flagellated marine bacteria 
\cite{thar2002conspicuous,thar2005complex}. Moreover, recent
computer simulations have confirmed crystallization \cite{bialke2012crystallization,redner2012structure,menzel2012soft} 
and proved that  melting 
of active crystals differs from its equilibrium counterpart.
However, though field-theoretical modelling of active systems
has been widely applied to orientational 
ordering phenomena \cite{ramaswamy2010mechanics,baskaran2008enhanced}, there is 
no such theory for translational ordering of active crystals nor has a systematic 
classification of active crystals been achieved.

Here we  present a microscopic field-theoretical approach to crystallization in 
active systems and we propose a minimal model which has
the necessary ingredients for both, crystallization and activity. In doing so, we combine  
the phase-field crystal model of freezing \cite{elder2002modeling} with the Toner-Tu model
for active systems \cite{toner1995long}  using the concept of 
dynamical density functional theory \cite{archer2004dynamical,wensink2008aggregation}. On the one hand, the phase-field-crystal 
(PFC) model as originally introduced by Elder and coworkers 
\cite{elder2002modeling,elder2004modeling} describes 
crystallization of passive particles on microscopic length and diffusive 
time scales. When brought into connection with dynamical density functional theory
\cite{elder2007phase,teeffelen2009derivation,tegze2009diffusion,jaatinen2009thermodynamics}, 
the PFC model represents in principle 
a microscopic theory for crystallization, and it has been successfully applied to a plethora of solidification phenomena 
\cite{elder2002modeling,elder2004modeling,stefanovic2006phase,stefanovic2009phase, chan2010plasticity,ramos2010dynamical,tegze2011faceting}.
On the other hand, Toner and Tu \cite{toner1995long} 
investigated the onset of collective motion in self-propelled systems 
from a general hydrodynamic point of view. Phenomenological 
coupling parameters of this model can in principle 
be justified by dynamical density functional theory \cite{wittkowski2011microscopic}, too, but 
it does not describe crystallization.

In our PFC model for active systems, we find a wealth of different crystallization 
phenomena. First, we identify two different types of active crystals
which we call ``resting'' and ``traveling'' depending on their averaged drift velocity.
A resting crystal possesses vanishing net particle flux whereas a traveling crystal is migrating with a nonzero
velocity while keeping its periodicity. Starting from a disordered initial state, 
a traveling crystal is typically formed  by a coarse-graining process of domains. 
The threshold in the driving strength upon which traveling crystals are formed depends
on the spontaneous local orientational order (as prescribed by the coupling parameters of the 
bare Toner-Tu model): if there is no such order, the threshold is 
finite, while there is no such threshold in the presence of spontaneous orientational order.
We further identify a transition from a hexagonal to a rhombic traveling crystal if 
the drive is increased further and find also resting and traveling lamellar phases
with one-dimensional periodic ordering. Finally the occurrence of honeycomb-like structures 
can be explained as well within our model. The knowledge and control over these crystalline states provides
an attractive starting point for the design of novel active materials since active crystals possess unique
structural, phononic, and rheological properties.

In the following, we first describe our model and then discuss 
numerical and analytical results. Our dynamical equations are for the local one-particle density
field $\psi_1(\mathbf{r},t)$ which is a conserved scalar order parameter and basically describes the reduced 
density modulation 
around a fixed averaged density $\bar{\psi}$ as in the 
traditional PFC model \cite{elder2002modeling,elder2004modeling}, and for
a polarization vector field $\mathbf{P}(\mathbf{r},t)$ which describes the local polar ordering. 
Activity enters into the equations via a nonzero self-propulsion velocity $v_0$.
In suitably scaled units of time, length and energy, our basic dynamic equations read
\begin{eqnarray}
\partial_{t}\psi_1 &=& \nabla^2\frac{\delta\mathcal{F}}{\delta\psi_1} - v_0\,\nabla\cdot\mathbf{P}, \label{eqpsi1}\\
\partial_{t}\mathbf{P} &=& \nabla^2\frac{\delta\mathcal{F}}{\delta\mathbf{P}} - D_r\frac{\delta\mathcal{F}}{\delta\mathbf{P}} - v_0\nabla\psi_1. 
\label{eqP}
\end{eqnarray}
Here, $D_r$ is the rotational diffusion coefficient of the particles
and  $\mathcal{F}$ is a free energy functional
of $\psi_1$ and $\mathbf{P}$ gained from density functional theory. 
The equations (\ref{eqpsi1}) and (\ref{eqP}) are consistent with phenomenological 
symmetry arguments and involve the simplest nontrivial coupling between the two order parameter fields
$\psi_1$ and $\mathbf{P}$. As outlined in the supplemental material \cite{supplemental}, they can also be derived from 
 microscopic dynamical density functional theory within an appropriate gradient and Taylor expansion
of the order parameter fields \cite{wensink2008aggregation,wittkowski2011polar}, see also \cite{cates2012motility}.
In the sequel, we shall consider two spatial dimensions only.

We now further specify the 
free energy functional $\mathcal{F}$ to $\mathcal{F}=\mathcal{F}_{pfc}+\mathcal{F}_{\mathbf{P}}$
where
\begin{equation}\label{Fpfc}
\mathcal{F}_{pfc} = \int d^2r\,\Big\{ \frac{1}{2}\psi\big[\varepsilon+(1+\nabla^2)^2\big]\psi + \frac{1}{4}\psi^4 \Big\}
\end{equation}
is the traditional PFC functional \cite{elder2002modeling,elder2004modeling} describing
 the tendency of the material to form periodic structures. Here, $\varepsilon$ sets the temperature \cite{elder2002modeling,elder2004modeling}, and the order parameter $\psi$ corresponds to the total density $\psi = \bar{\psi}+\psi_1$. 
The polarization-dependent part 
\begin{equation}\label{FP}
\mathcal{F}_{\mathbf{P}}=\int d^2r\,\Big\{ \frac{1}{2}C_1\mathbf{P}^2+\frac{1}{4}C_4(\mathbf{P}^2)^2 \Big\}
\end{equation} 
 describes local orientational ordering due to the active driving 
following  the approach by Toner and Tu \cite{toner1995long} for neglected convection. 
The functional possesses two coupling parameters $C_{1}$ and $C_{4}$ which govern the local
 orientational ordering due to the drive.
If $C_1=C_4=0$, only gradients in the density $\psi_1$ can induce local polar order $\mathbf{P}$ of the active driving. For $C_1>0$ ($C_4=0$) diffusion tends to reduce the polar order generated by the density gradients. In the third case, $C_1<0$ and $C_4>0$, a net local driving  spontaneously emerges already in the absence of density gradients. 

Clearly, on the one hand, for vanishing self-propulsion $v_0=0$, Eqs.~(\ref{eqpsi1}) and (\ref{eqP}) decouple and the density equation reduces to the usual phase field crystal model \cite{elder2002modeling,elder2004modeling}. On the other hand, if $\mathcal{F}_{pfc}$ is neglected, the remaining terms are contained in the model by Toner et al.~\cite{toner1995long,toner2005hydrodynamics}, except for the higher-order term in $\mathbf{P}$ that contributes to translational diffusion. 
Summarizing, Eqs.~(\ref{eqpsi1})--(\ref{FP}) form a minimal approach
 to characterize crystallization in actively driven systems.

We numerically determined the phase diagram by scanning the $\bar{\psi}$-$\varepsilon$ plane while keeping the parameters $C_1$, $C_4$, and $v_0$ fixed.
As for any numerical result reported subsequently, we proceeded in the following way. For each set of parameter values $(\bar{\psi},\varepsilon,C_1,C_4,v_0)$ we started from random initial conditions and then iterated Eqs.~(\ref{eqpsi1})--(\ref{FP}) forward in time. Numerical measurements were carried out after equilibration, and a systematic finite size study was performed to test the validity of our results. 

\begin{figure}
\centerline{\includegraphics[width=8.5cm]{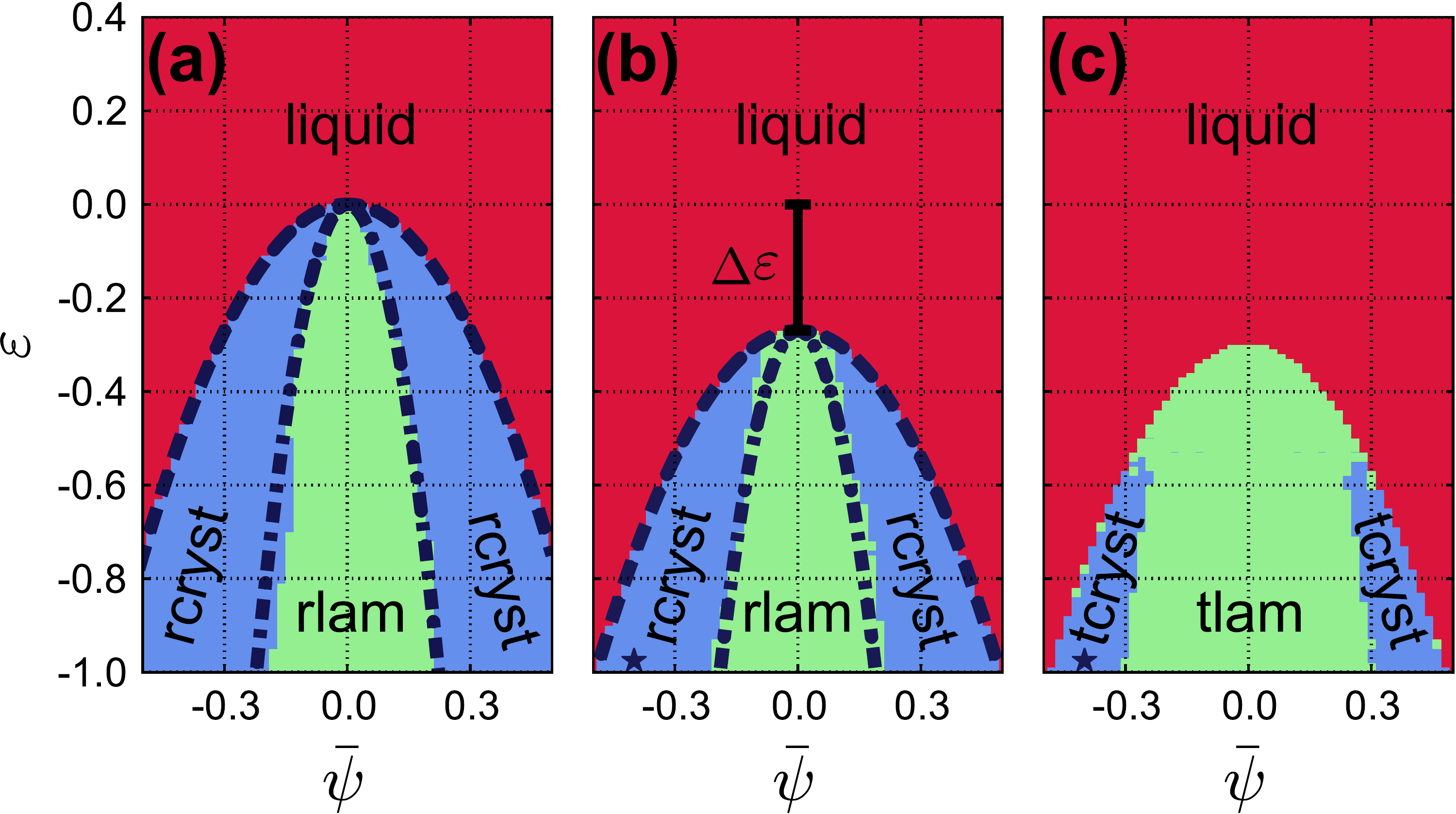}}
\caption{Phase diagrams (``rcryst'': resting crystals; ``rlam'': resting lamellae; ``tcryst'': traveling crystals; ``tlam'': traveling lamellae). (a) For $C_1=0$, $v_0=0$ the equilibrium phase field crystal model is recovered. Equilibrium phase boundaries given by the energy functional are indicated for the liquid--hexagonal (dashed line) and hexagonal--lamellar (dash-dotted line) transitions. (b) For $C_1=0.2$, $v_0=0.35$ the structures are still at rest, but the phase boundaries are shifted by a value $\Delta\varepsilon$. (c) For $C_1=0.2$, $v_0=0.7$ the structures are traveling and phase boundary lines are omitted for clarity. The black stars in the bottom left of panels (b) and (c) mark the intersection points with the curve in Fig.~\ref{Fparup}. In all cases $C_4=0$, $D_r=0.5$.}
\label{phasediagram}
\end{figure}

For the decoupled case $v_0=0$, the equilibrium phase diagram \cite{elder2004modeling} corresponding to the energy functional Eq.~(\ref{Fpfc}) is shown in Fig.~\ref{phasediagram}(a). 
For nonzero active drive $v_0$, we will first report on the case $C_1>0$. 

When we moderately increase $v_0$ from zero for $C_1>0$, the phase boundaries undergo a temperature shift $\Delta\varepsilon$ to lower temperatures. An example is depicted in Fig.~\ref{phasediagram}(b). Comparison to Fig.~\ref{phasediagram}(a) shows that switching on the active drive melts crystals and lamellae close to the liquid phase boundary. The patterns still remain at rest, however. 
For this case, a linear stability analysis and derived amplitude equations for $\psi_1$ and $\mathbf{P}$ predict $\Delta\varepsilon\propto v_0^2/C_1$, which was also verified numerically. 

\begin{figure}
\centerline{\includegraphics[width=8.5cm]{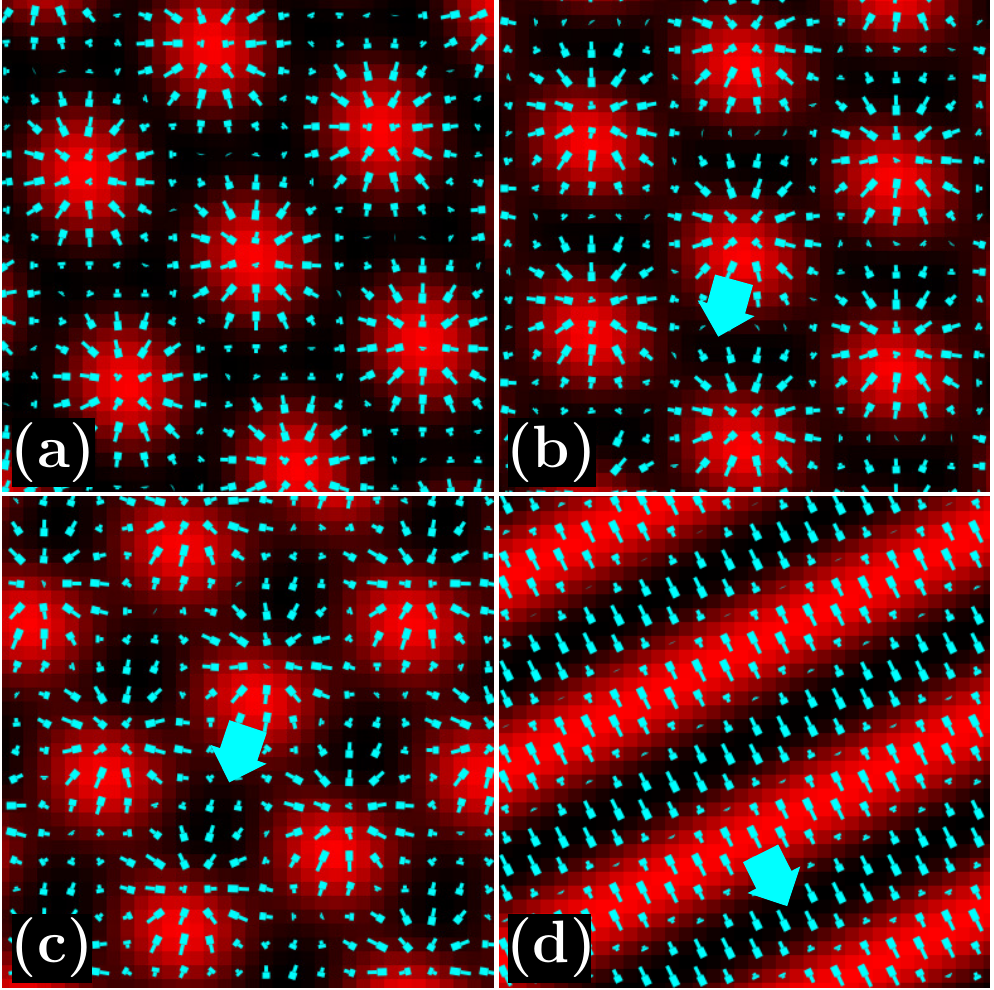}}
\caption{Snapshots of the order parameter fields for different phases that are observed when increasing the active drive $v_0$ at $(\bar{\psi},\varepsilon,C_1,C_4)=(-0.4,-0.98,0.2,0)$: (a) resting hexagonal, $v_0=0.1$, (b) traveling hexagonal, $v_0=0.5$, (c) traveling quadratic, $v_0=1$, (d) traveling lamellar, $v_0=1.9$. The phases are depicted by plotting the density field $\psi_1$: brighter color corresponds to higher densities. Thin bright needles illustrate the polarization field $\mathbf{P}$ that points from the thick to the thin ends. In panels (b)--(d) the predominant direction of motion is indicated by the bright arrows. Only a fraction of the numerical calculation box is shown.}
\label{snapshots}
\end{figure}

We present an example snapshot of the resting crystalline phase in Fig.~\ref{snapshots}(a). 
The peaks of the density distribution $\psi_1$ form a hexagonal lattice as dictated by the PFC energy functional. $\mathbf{P}$ points down the density gradients. Consequently the polarization field forms ``$+1$''-defects centered at the density peaks. 
Since $\mathbf{P}$ describes the local direction of active drive, density is convected out of the peaks by the active propulsion $v_0$. This mechanism counteracts the density diffusion into the peaks described by the PFC energy functional. Therefore lower temperatures are necessary for the patterns to form in the presence of an active drive, corresponding to the temperature shift $\Delta\varepsilon$ in Fig.~\ref{phasediagram}(b). In the resting crystalline and lamellar case, both tendencies balance each other so that the averaged net particle flux vanishes. 

When we increase the active drive, we find that the density peaks start to travel above a critical value $v_{0,c}$. Such a state is illustrated in Fig.~\ref{snapshots}(b). As we can see, the centers of the density peaks are now shifted with respect to centers of the ``$+1$''-defects in the polarization field. This reduced symmetry induces active propulsion: a net orientation of the polarization field emerges when averaged over the area of a single density peak. The consequence is an active convection of each density peak, originating from the PFC density modulation. These results are in agreement with a linear stability analysis of Eqs.~(\ref{eqpsi1}) and (\ref{eqP}) which predicts that propagating modes appear above a threshold value $v_{0,c}$. 

With further increasing $v_0$ the hexagonal pattern is deformed to a rhombic one. In the end, we observe a nearly quadratic structure as depicted in Fig.~\ref{snapshots}(c). This structural hexagonal--rhombic--quadratic transition appears to be smooth and continuous. 

Finally, we observe that the traveling crystal can be melted into a traveling lamellar state if $v_0$ is increased to still higher values. An example snapshot of such traveling lamellae is shown in Fig.~\ref{snapshots}(d). In contrast to the hexagonal--rhombic--quadratic distortion of the traveling crystalline lattices, the traveling crystalline--lamellar transition 
occurs rather abruptly. The transition is also evident when we compare the two phase diagrams in Figs.~\ref{phasediagram}(b) and \ref{phasediagram}(c). There, with increasing $v_0$, the traveling lamellar regions grow into the traveling crystalline regions.

To quantify the scenario further, we tracked the motion of each density peak. We determined the individual peak velocities $\mathbf{v}_i$, where $i=1,..,N_p$ and $N_p$ denotes the number of peaks. Samples of up to $1000$ density peaks were investigated. The sample-averaged peak velocity magnitude follows as $v_m=\sum_{i=1}^{N_p} \|\mathbf{v}_i\|/{N_p}$. In addition, we calculated the degree of polar orientational order of the normalized peak velocities $p_{v}=\big\|(\sum_{i=1}^{N_p} \mathbf{v}_i/\|\mathbf{v}_i\|)\big\|/{N_p}$. This order parameter detects whether the peaks move coherently (collectively) into the same direction. 
\begin{figure}
\centerline{\includegraphics[width=8.5cm]{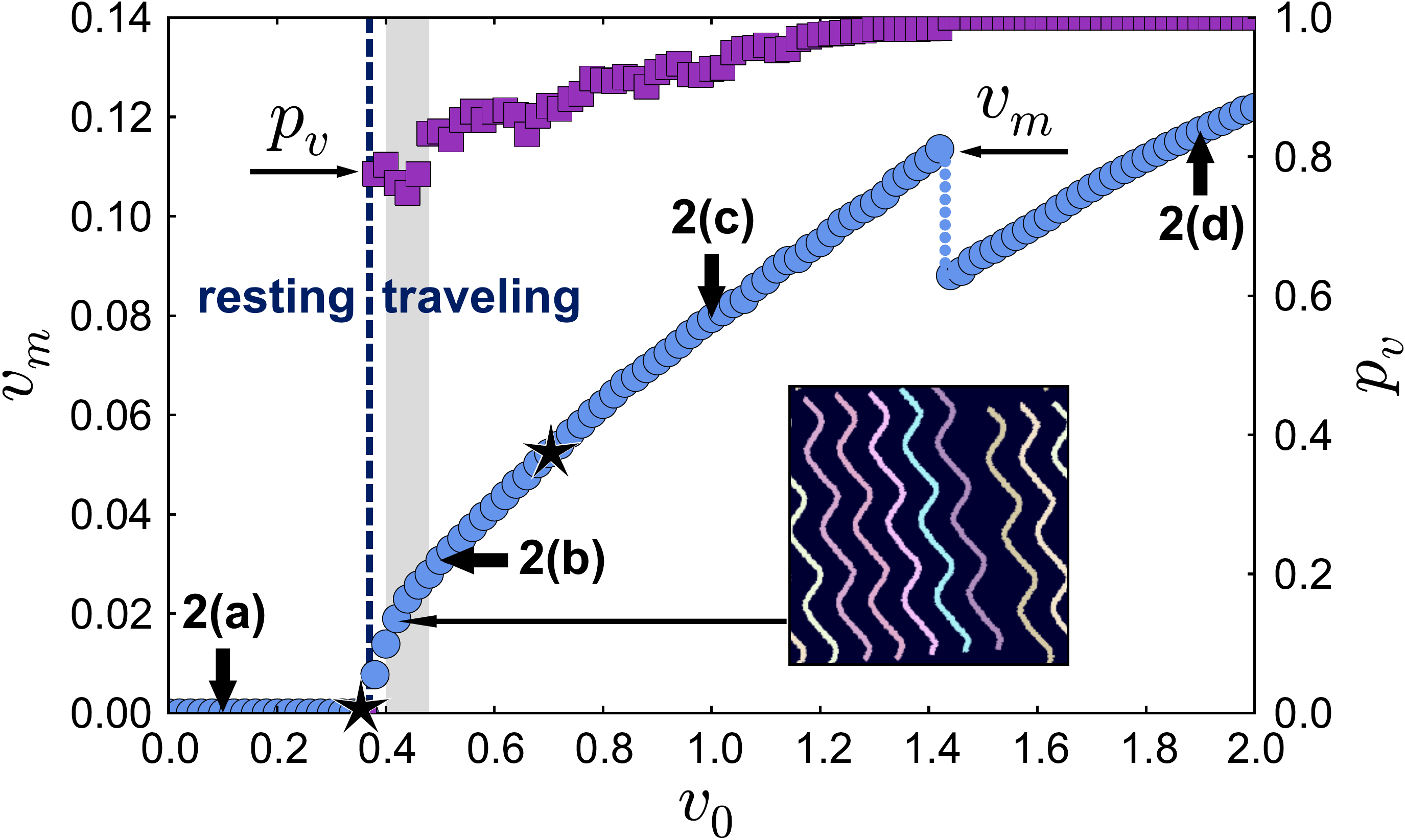}}
\caption{Sample-averaged magnitude $v_m$ of the crystal peak velocities (left scale) and polar order parameter $p_v$ of the crystal peak velocity vectors (right scale) as a function of $v_0$ for $(\bar{\psi},\varepsilon,C_1,C_4)=(-0.4,-0.98,0.2,0)$. The threshold corresponds to the onset of collective crystalline motion. 
Thick arrows mark the positions where the snapshots of Fig.~\ref{snapshots} were taken; the black stars indicate the intersection points with the phase diagrams in Figs.~\ref{phasediagram}(b) and \ref{phasediagram}(c). The region above threshold where regular swinging motion could be observed is marked in gray. 
Inset: 
peak trajectories illustrating a state of regular swinging motion in a hexagonal crystal; different colors correspond to different peaks; only trajectories of a horizontal row of density peaks are shown that started at the bottom and were traveling to the top of the picture while tracking was performed. 
}
\label{Fparup}
\end{figure}

For $C_1>0$, Fig.~\ref{Fparup} clearly illustrates the existence of a threshold value $v_{0,c}$ at which propagation starts. As indicated in the inset, we observed a regular {\em swinging} motion of the hexagonal crystals close to the threshold. 
With increasing values of active drive, we subsequently find the states illustrated in Fig.~\ref{snapshots}(a)--(d). The averaged peak velocity magnitude $v_m$ monotonously increases, until it abruptly drops at the transition from traveling quadratic crystals to traveling lamellae. We can obtain a traveling quadratic crystal from superimposing perpendicularly oriented and traveling lamellae. Their intersections form peaks that travel $\sqrt{2}$ times faster than each single lamella by itself, which approximately corresponds to the magnitude of the drop in the $v_m$-curve.

\begin{figure}
\centerline{\includegraphics[width=8.5cm]{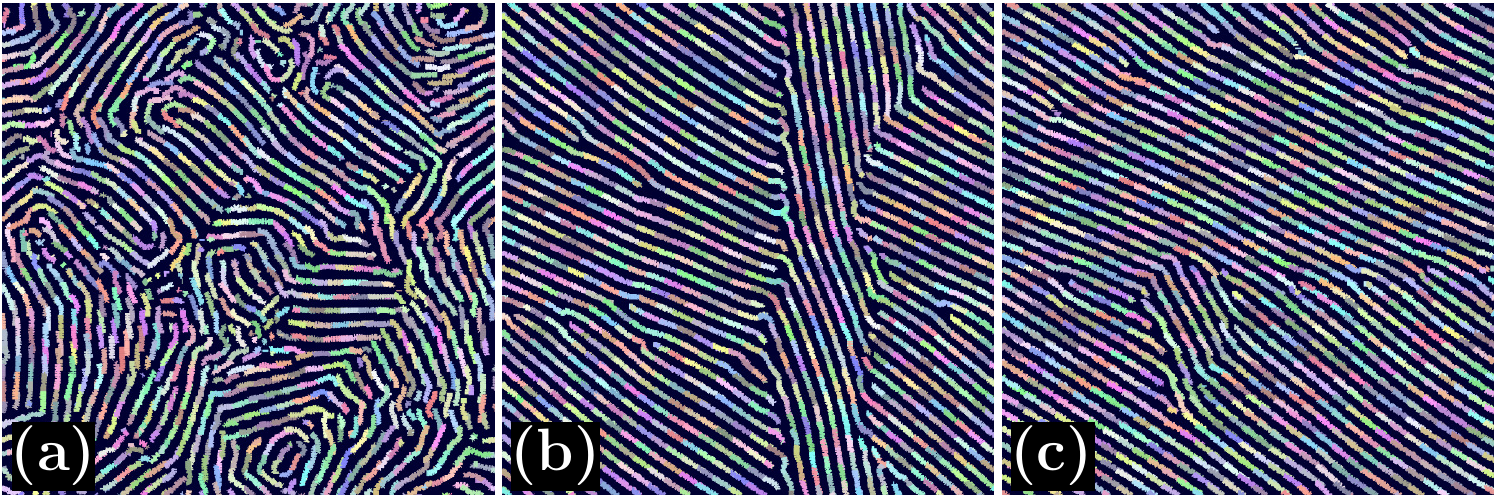}}
\caption{Coarse-graining in a sample of about 1000 traveling density peaks. The sample is in the traveling hexagonal crystalline state. Depicted are instant pieces of all peak trajectories, drawn in different colors. These pieces connect to lines of equal orientation within collectively moving domains. Over time, some domains of equal velocity orientation grow on the cost of others. Parameter values correspond to Fig.~\ref{snapshots}(b). Dimensionless times \cite{supplemental}: (a) 4500, (b) 50000, (c) 70000.}
\label{coarsegrain}
\end{figure}

Furthermore, we observe in Fig.~\ref{Fparup} that the polar peak velocity order parameter $p_v$ jumps to a value close to one at the threshold and then further increases. This indicates that after equilibration of the sample the density peaks migrate coherently into the same direction and the crystal travels as a single object. However, at each value of $v_0$, this collective motion has to first develop from the disordered initial state. The latter process occurs through a coarse-graining dynamics as qualitatively illustrated in Fig.~\ref{coarsegrain}.
The panels depict traveling hexagonal peak trajectories at different times of coarse-graining. 
First, collectively moving crystalline domains form from the disordered initial state. 
However, the migration directions of different domains are not identical. 
Over time, some domains grow on the cost of others, until a collectively traveling crystal emerges.

Finally, if we set $C_1<0$ and $C_4>0$, a net polar direction $\mathbf{P}$ of self-propulsion spontaneously occurs as in the Toner-Tu model \cite{toner1995long,toner2005hydrodynamics}. 
For this scenario we never observed a finite threshold value of $v_0$. Propagating structures evolved for all tested nonzero values of $v_0$. Again, a transition from hexagonal to rhombic to quadratic and then to lamellar structures was observed with increasing $v_0$. 
Furthermore, we note that our equations of motion are invariant under the transformation $\bar{\psi}\rightarrow -\bar{\psi}$, $\psi_1\rightarrow -\psi_1$, $\mathbf{P}\rightarrow -\mathbf{P}$. Because of these symmetry relations, our analysis equally applies for the investigation of active honeycomb textures that follow from $\bar{\psi}>0$. Such textures were observed for flagellated marine bacteria \cite{thar2002conspicuous,thar2005complex}.

In summary, we extended the phase field crystal model \cite{elder2002modeling,elder2004modeling} to 
active systems by combining it with the approach of 
Toner et al.~\cite{toner1995long,toner2005hydrodynamics}. 
As a result,
the active drive favors the liquid and lamellar states in the PFC phase diagram
and induces a wealth of new active crystalline states of hexagonal,
honeycomb, rhombic, and quadratic texture. The global motion of all these
  structured states is  either ``resting'' or ``traveling''. The transition
from ``resting'' to ``traveling'' involves a complex 
intermediate swinging motion. When prepared from an initially
disordered state, traveling crystals emerge through coarse-graining
from a multidomain texture.

Our model can be extended from two to three spatial dimensions where more crystalline lattice structures 
become stable \cite{jaatinen2009thermodynamics} and to binary mixtures of driven and undriven particles promising a rich variety of mixed active crystals. 
 In principle, our predictions are verifiable in experiments on self-propelled particles 
and in particle-resolved computer simulations at high density \cite{bialke2012crystallization,redner2012structure,menzel2012soft}. 
Very recently, traveling crystals have in fact been found in such simulations \cite{frey2012}.  
Since the new traveling crystalline structures show a nontrivial dynamical response, 
they may serve as a building block for a new class of active matter with 
unusual rheological, phononic, and possibly also photonic properties.

\begin{acknowledgments}
The authors thank Erwin Frey, Takao Ohta, and Raphael Wittkowski for stimulating discussions. Support from the Deutsche Forschungsgemeinschaft through SFB TR6, SPP 1296, and the German--Japanese project LO 418/15 is gratefully acknowledged. 
\end{acknowledgments}

\end{document}